\documentstyle[preprint,aps]{revtex}
\tighten
\begin{document}
\draft
\preprint{\vbox{Submitted to Physical Review C}}
\title{Coherent \boldmath{$\!\!\eta$}-photoproduction from nuclei \\
       in a relativistic impulse approximation approach}
\author{J. Piekarewicz$^{1}$, A.J. Sarty$^{2}$, and 
        M. Benmerrouche$^{3}$}
\address{${}^{1}$Supercomputer Computations Research Institute, \\
                 Florida State University, 
	         Tallahassee, FL 32306, USA}
\address{${}^{2}$Department of Physics, 
                 Florida State University, 
                 Tallahassee, FL 32306, USA}
\address{${}^{3}$Saskatchewan Accelerator Laboratory,
	         University of Saskatchewan, \\
		 Saskatoon, SK~S7N~5C6, Canada}
\date{\today}
\maketitle
 
\begin{abstract}
  We study coherent $\eta$-photoproduction from nuclei in a
relativistic impulse approximation approach. For the elementary
production amplitude we use a standard relativistic parameterization
based on a set of four Lorentz- and gauge-invariant amplitudes. The
photonuclear amplitude is evaluated without recourse to a
nonrelativistic reduction; the full relativistic structure of the
amplitude is maintained. On general arguments we show that the coherent
process is sensitive to only one of the elementary amplitudes.
Moreover, we show that the nuclear structure information is fully
contained in the ground-state tensor density. The tensor density is
evaluated in a mean-field approximation to the Walecka model and it is
shown to be sensitive to relativistic effects. Distortion effects are
incorporated through an $\eta$-nucleus optical potential that is
computed in a simple ``$t\rho$'' approximation.
\end{abstract}

\narrowtext
 
\section{Introduction}
\label{sec:intro}
 The coherent photoproduction of pseudoscalar mesons (such as
$\pi^{0}$ and $\eta$) offers a unique opportunity for the
investigation of nucleon-resonance formation and propagation through 
the nuclear medium. The advent of more powerful and sophisticated 
machines---such as TJNAF and MAMI---will challenge, now more than ever, 
our theoretical understanding of this fundamental process.

 Although experimentally challenging, the
coherent photoproduction reaction offers numerous advantages. Because
the nucleus remains in its ground state, all nucleons participate
coherently in the reaction leading to enhanced cross sections relative
to incoherent processes to discrete nuclear states. Theoretically, the
coherent process acts as a spin-isospin filter by selecting a
particular (scalar-isoscalar) component of the elementary
photoproduction amplitude. In this way the reaction can be used to
discriminate between various theoretical models that provide an
equally good description of the elementary process. Moreover, all
nuclear structure information is contained in, at most, a few
ground-state densities. Indeed, nonrelativistic
plane-wave-impulse-approximation analyses suggest that the
photonuclear amplitude is directly proportional to the isoscalar (or
matter) density~\cite{bofmir86,cek87,bentan90,ndu91,tryfik94}.  Since
the ground-state charge density is very accurately determined from
electron-scattering experiments, the coherent reaction becomes---after
making some plausible assumptions about the neutron density---an ideal
tool in the determination of the scalar-isoscalar component of the
elementary amplitude, and its possible modification in the many-body
environment. Finally, because the coherent process is sensitive to the
whole nuclear volume, one can place stringent limits on the form of
the meson-nucleus optical potential. This unique character of the
coherent reaction has already led to an experimental effort to
measure coherent pion photoproduction from ${}^{4}$He~\cite{tieger84}
and ${}^{12}$C~\cite{gothe95}, and coherent eta photoproduction
from ${}^{4}$He~\cite{ahrens93} up to photon energies of about
800 MeV; possibilities for extensions to higher energies, and other
nuclei, exist both at the Bonn ELSA facility and at TJNAF.

 Most theoretical analyses of the elementary process start with a
model-independent parameterization of the photoproduction amplitude in
terms of four Lorentz- and gauge-invariant amplitudes~\cite{cgln57}.
It is then customary to evaluate this amplitude between on-shell
nucleon spinors, thereby leading to the well known CGLN form for the
photoproduction operator in terms of Pauli---rather than
Dirac---spinors~\cite{cgln57}. For the the calculation of the
photonuclear reaction one usually adopts the impulse
approximation~\cite{bofmir86,cek87,bentan90,ndu91,tryfik94}; one
assumes that the elementary (on-shell) amplitude is not modified in
the many-body environment. For closed-shell (spin-saturated) nuclei
the photonuclear process then becomes, in the plane-wave limit, a
simple product of the elementary scalar-isoscalar amplitude times the
Fourier transform of the ground-state matter density. One then
improves on the plane-wave description by incorporating distortions 
into the propagation of the outgoing meson.

 In this paper we propose to carry out the above theoretical 
program---without recourse to a nonrelativistic reduction of 
the elementary amplitude. The main difference relative to the 
standard nonrelativistic approach stems from the fact that in 
the present relativistic framework the lower components of the 
nucleon spinors will be determined dynamically, rather than 
from the free-space relation. The nuclear structure information
will be contained in a few ground-state densities that will be
computed using a mean-field approximation to the Walecka 
model~\cite{serwal86}. It is worth mentioning that the Walecka 
model has enjoyed considerable success in describing ground-state 
properties of many nuclei and, indeed, rivals some of the best 
available nonrelativistic calculations.

\section{Formalism}
\label{sec:formal}
The differential cross section in the center-of-momentum frame
(cm) for the coherent photoproduction of $\eta$-mesons is given 
by~\cite{bentan90} 
%%%
\begin{equation}
   \left({d\sigma \over d\Omega}\right)_{\rm cm} =
   \left({M_{\lower 1pt \hbox{$\scriptstyle T$}} 
    \over 4\pi W}\right)^{2} 
   \left({q_{\rm cm} \over k_{\rm cm}}\right)
   {1 \over 2}\sum_{\lambda} 
   |T_{\lower 2pt \hbox{$\scriptstyle\lambda$}}|^2 \;,
 \label{dsigmaa}
\end{equation}
%%%
where $M_{\lower 1pt \hbox{$\scriptstyle T$}}$ is the mass of the
target nucleus, $W$ is the total energy in the cm-frame, while 
$k_{\rm cm}$ and $q_{\rm cm}$ are the three-momentum of the photon 
and $\eta$-meson in the cm-frame, respectively. Note that the formalism 
is independent of the mass of the produced meson so it can be applied
without modification to the coherent photoproduction of any
pseudoscalar meson. The relativistic-invariant amplitude for the
coherent photoproduction from ($J^{\pi}\!=\!0^{+};T\!=\!0$) nuclei can
be written in the following way:
%%%
\begin{equation}
 T_{\lower 2pt \hbox{$\scriptstyle\lambda$}} =
 \epsilon_{\mu}(\hat{\bf k},\lambda) 
 \langle A(p'); \eta(q) | J^{\mu} | A(p) \rangle \;,
 \label{tampl}
\end{equation}
%%%
where $\epsilon_{\mu}(\hat{\bf k},\lambda)$ is the polarization
vector of the photon which couples to a conserved electromagnetic 
current given by:
%%%
\begin{equation}
 \langle A(p'); \eta(q) | J^{\mu} | A(p) \rangle \equiv
 \varepsilon^{\mu\nu\alpha\beta} k_{\nu}q_{\alpha}p_{\beta}
 {1 \over W} F_{0}(s,t) \;.
 \label{emcurrent}
\end{equation}
%%%
Here $p(p'\!=\!p\!+\!k\!-\!q)$ is the four momentum of the 
initial(final) nucleus and $\varepsilon^{\mu\nu\alpha\beta}$
is the relativistic Levi-Civita symbol 
($\varepsilon^{0123}\equiv-1$). Note that all the dynamical 
information about the coherent process is contained in a 
single Lorentz-invariant form factor $F_{0}(s,t)$, which 
depends on the Mandelstam variables $s=(k+p)^{2}$ and 
$t=(k-q)^{2}$. One can now carry out the appropriate algebraic 
manipulations to obtain the following---model-independent---form
for the coherent photoproduction cross section:
%%%
\begin{equation}
   \left({d\sigma \over d\Omega}\right)_{\rm cm} =
   \left({M_{\lower 1pt \hbox{$\scriptstyle T$}} 
    \over 4\pi W}\right)^{2} 
   \left({q_{\rm cm} \over k_{\rm cm}}\right)
   \left({1 \over 2}k_{\rm cm}^{2}q_{\rm cm}^{2}
   \sin^{2}\theta_{\rm cm}\right)|F_{0}(s,t)|^{2} \;,
 \label{dsigmab}
\end{equation}
%%%
where $\theta_{\rm cm}$ is the scattering angle in the cm-frame.

 We now proceed to compute the Lorentz invariant form factor
in a relativistic impulse approximation. For the elementary
$\gamma N \rightarrow \eta N$ amplitude we use a model-independent
parameterization given in terms of four Lorentz- and gauge-invariant 
amplitudes:
%%%
\begin{equation}
   T(\gamma N \rightarrow \eta N) =
   \sum_{i=1}^{4} A_{i}(s,t) {M}_{i} \;.
 \label{telema}
\end{equation}
%%%
For the invariant matrices we use the following standard 
set~\cite{bentan90}:
%%%
\begin{mathletters}
 \begin{eqnarray}
  {M}_{1} &=& -\gamma^{5}\rlap/{\epsilon}\;\rlap/{k} \;, \\
  {M}_{2} &=& 2\gamma^{5}
    \Big[(\epsilon \cdot p)(k \cdot p') -
         (\epsilon \cdot p')(k \cdot p) \Big]  \;, \\
  {M}_{3} &=& \gamma^{5}
    \Big[\rlap/{\epsilon}\,(k \cdot p) -
         \rlap/{k}(\epsilon \cdot p) \Big]     \;, \\
  {M}_{4} &=& \gamma^{5}
    \Big[\rlap/{\epsilon}\,(k \cdot p') -
         \rlap/{k}(\epsilon \cdot p') \Big]    \;. 
  \label{allm}
 \end{eqnarray}
\end{mathletters}
%%%
Note that, although standard, this is only one particular form 
for the elementary amplitude. Many other choices---all of them 
equivalent on shell---are possible~\cite{ndu91,cgln57}. However, 
in the absence of a detailed microscopic model of the elementary 
process it becomes ambiguous on how to take the elementary amplitude 
off shell. The above form of the elementary amplitude, or one very 
close to it, is ubiquitous in theoretical studies, so we adopt it 
here as well.

 To proceed, we perform some simple algebraic manipulations on the
elementary amplitude so that the parity and Lorentz transformation
properties of the various bilinear covariants become manifest. That 
is,
%%%
\begin{equation}
   T(\gamma N \rightarrow \eta N) = 
      \Big(F_{T}^{\alpha\beta}\sigma_{\alpha\beta} +
           F_{P}\gamma_{5} +
           F_{A}^{\alpha}\gamma_{\alpha}\gamma_{5}\Big)\;,
 \label{telemb}
\end{equation}
%%%
where tensor, pseudoscalar, and axial-vector amplitudes have
been introduced
%%%
\begin{mathletters}
 \begin{eqnarray}
   &&F_{T}^{\alpha\beta} = 
   {1 \over 2}\varepsilon^{\mu\nu\alpha\beta}\,
    \epsilon_{\mu}\,k_{\nu}\,A_{1}(s,t) \;, \\
   &&F_{P} = 
    2 \Big[(\epsilon \cdot p)(k \cdot p') -
           (\epsilon \cdot p')(k \cdot p) \Big]A_{2}(s,t)  \;, \\
   &&F_{A}^{\alpha} = 
    \Big[(\epsilon \cdot p) k^{\alpha} -
         (k \cdot p) \epsilon^{\alpha}\Big]A_{3}(s,t) +
    \Big[(\epsilon \cdot p')k^{\alpha}-
         (k \cdot p')\epsilon^{\alpha}\Big]A_{4}(s,t) \;.
  \label{allf}
 \end{eqnarray}
\end{mathletters}
%%%
Note that for this particular form of the elementary amplitude 
no scalar nor vector invariants appear. For closed-shell nuclei 
an enormous simplification ensues, as a result of the pseudoscalar
and axial-vector ground-state densities being identically zero. 
This implies that the coherent reaction is sensitive to only 
the $A_{1}$ component of the elementary amplitude. Moreover, 
all the nuclear-structure information is contained in the 
ground-state tensor density~\cite{serwal86}. This is in contrast 
to nonrelativistic approaches in which the coherent amplitude is 
proportional to the conserved vector 
density~\cite{bofmir86,cek87,bentan90,ndu91,tryfik94}. 
Thus, in a relativistic plane-wave impulse approximation, the 
Lorentz-invariant form factor acquires a remarkable simple form:
%%%
\begin{equation}
  F_{0}^{\scriptscriptstyle PW}(s,t) = i A_{1}(\tilde{s},t) 
   {\rho_{\lower 3pt \hbox{$\scriptstyle T$}}(Q) / Q} \;.
  \label{fpwia}
\end{equation} 
%%%  
Note that $\tilde{s}$ represents the effective (or optimal) value 
of the Mandelstam variable $s$ at which the elementary amplitude 
should be evaluated~\cite{cek87} and 
$Q \equiv |{\bf k}_{\rm cm}-{\bf q}_{\rm cm}|\simeq \sqrt{-t}$.

 The elementary $\eta$-production amplitude used is constructed in
an effective Lagrangian approach as detailed in 
references~\cite{benm92,bmz95}. The dynamical content of the amplitude
consists of: i) s- and u-channel nucleon Born terms, ii) t-channel 
vector-meson exchange, and iii) intermediate excitation of spin-1/2 
and spin-3/2 resonances in the s- and u-channels (resonating only in 
the s-channel). The free parameters in the elementary model were fixed
by achieving good fits to the latest $p(\gamma,\eta)p$~\cite{krus95a} 
and $d(\gamma,\eta )pn$~\cite{krus95b} data. The inclusion of two 
intermediate resonances, the $S_{11}$(1535) and $D_{13}$(1520),
generates good fits. As is well known, the $S_{11}$(1535) clearly
dominates the elementary reaction, but the $D_{13}$(1520) 
contribution is needed to reproduce the measured angular
distributions. The model parameters we use here are identical to 
those determined in Ref.~\cite{bmz95}. Fig.~\ref{figzero} shows a 
sample comparison of this elementary model to some of the data from 
Ref.'s~\cite{krus95a} and~\cite{krus95b}; the data for the elementary 
neutron amplitudes are from Ref.~\cite{krus95b}, as extracted from the
measured $d( \gamma,\eta )np$ data.  Good agreement is found between the 
elementary model and the data for all energies and angles measured, 
and for both the proton and extracted neutron amplitudes. Also shown 
in Fig.~\ref{figzero} is the contribution from the $S_{11}$(1535) 
excitation term only.  Clearly, intermediate $S_{11}$(1535) resonance 
excitation dominates the production amplitude, but is not sufficient 
on its own to adequately describe the data. Note that, when embedded 
into the impulse-approximation calculation for coherent production from 
closed-shell nuclei, intermediate $S_{11}$(1535) excitation is 
suppressed due to spin-isospin considerations, and the role of the 
$D_{13}$(1520) is enhanced.  Indeed, it is just this suppression
of the dominant s-wave term---allowing enhancement of the 
non-dominant contributions---which initially provoked interest in 
this coherent reaction~\cite{bentan90}

 The ground-state tensor density is evaluated in a mean-field
approximation to the Walecka model. For closed-shell nuclei the 
relativistic spinors can be classified according to a generalized 
angular momentum $\kappa$ and can be written in a two-component 
representation; i.e.,
%%%
\begin{equation}
  {\cal U}_{\alpha}({\bf x}) = {1 \over r}
   \left[
    \begin{array}{c}
      \phantom{i}
       g_{a}(r)
      {\cal{Y}}_{+\kappa m}(\hat{\bf{r}})    \\
       i
       f_{a}(r)
      {\cal{Y}}_{-\kappa m}(\hat{\bf{r}})    
     \end{array}
   \right] \;; \quad \Big(\alpha \equiv (a;m) = (n,\kappa;m)\Big) \;,
 \label{mfspinors}
\end{equation}
%%%
where the upper and lower components are expressed in terms of 
spin-spherical harmonics defined by
%%%
\begin{equation}
  {\cal{Y}}_{\kappa m}(\hat{\bf{r}}) \equiv
  \langle\hat{\bf{r}}|l{\scriptstyle{1\over 2}}jm\rangle \;; \quad
  j=|\kappa|-{1\over 2} \;; \quad
  l=\cases{   \kappa\;,   & if $\kappa > 0$; \cr
           -1-\kappa\;,   & if $\kappa < 0$. \cr}
 \label{curlyy}
\end{equation}
%%%
For spin-saturated nuclei only three ground-state densities do not
vanish~\cite{serwal86}; these are the scalar and timelike-vector 
densities---used to compute the mean-field ground state---and the 
tensor density defined by
%%%
\begin{mathletters}
 \begin{eqnarray}
  \Big[
    \rho_{\lower 3pt \hbox{$\scriptstyle T$}}(r)\,\hat{r}
  \Big]^{i} &=& 
  \sum_{\alpha}^{\rm occ}
   \overline{{\cal U}}_{\alpha}({\bf x})\,
   \sigma^{{\scriptscriptstyle 0}i} \,
              {\cal U}_{\alpha}({\bf x}) \;, \\
   \rho_{\lower 3pt \hbox{$\scriptstyle T$}}(r) &=&
   \sum_{a}^{\rm occ}
   \left({2j_{a}+1 \over 4\pi r^{2}}\right)
   2g_{a}(r)f_{a}(r) \;.
 \label{rhotr}
 \end{eqnarray}
\end{mathletters}
%%%
For the coherent reaction it is only the Fourier transform of the
latter that is needed
%%%
\begin{equation}
    \rho_{\lower 3pt \hbox{$\scriptstyle T$}}(Q) =
     4\pi \int_{0}^{\infty} dr \, r^2
     j_{\lower 2pt \hbox{$\scriptstyle 1$}}(Qr)
    \rho_{\lower 3pt \hbox{$\scriptstyle T$}}(r) \;.
 \label{rhotq}
\end{equation}
%%%
Note that the tensor density is linear in the lower (or small) 
component of the single-particle wave function; this is in
contrast to the scalar and vector densities where the lower 
component enters as an $(f/g)^{2}$ correction. Thus, the tensor 
density is small and not well constrained by experiment. Yet, 
it is as fundamental as the vector density measured in electron 
scattering. Moreover, the tensor density is interesting because 
it is sensitive to the relativistic components of the wave 
function. Indeed, the mean-field approximation to the Walecka 
model is characterized by the existence of large Lorentz scalar
and vector potentials that are responsible for a substantial
enhancement of the lower components of the single-particle
wave functions. This enhancement is at the heart of the
phenomenological success enjoyed by the Walecka model. Thus,
the enhancement of the tensor density (to be shown later) 
represents an inescapable prediction of the model. 

 Although simple and illuminating, the plane-wave formalism 
must be modified so that final-state interactions between the
outgoing meson and the nucleus can be incorporated. This is
done via an $\eta$-nucleus optical potential of the $t\rho$ 
form:
%%%
\begin{equation}
  2\omega_{\eta}V(r) = - b 
  \rho_{\lower 3pt \hbox{$\scriptstyle V$}}(r) \;.
 \label{trho}
\end{equation}
%%%
Here $\rho_{\lower 3pt \hbox{$\scriptstyle V$}}(r)$
is the conserved vector density---also computed in the 
mean-field approximation to the Walecka model---and $b$ 
is a two-body ($\eta N$) parameter taken from 
Ref.~\cite{bentan90} and parameterized in the following way:
%%%
\begin{mathletters}
 \begin{eqnarray}
  b(p_{\rm lab}) & \equiv & 
  (\alpha + \beta p_{\rm lab} + \gamma p_{\rm lab}^{2})^{-1} \;, \\
   \alpha &=& (+0.136,-0.052)\,{\rm fm}^{-1} \;, \\
    \beta &=& (+0.035,-0.072)                \;, \\
   \gamma &=& (-0.061,+0.009)\,{\rm fm}      \;.
 \label{bfit}
 \end{eqnarray}
\end{mathletters}
%%%
The Klein-Gordon equation for the outgoing $\eta$-meson can 
now be solved in each angular-momentum channel resulting in 
a coherent form factor expressed in a partial-wave series. 
That is,
%%%
\begin{equation}
    F_{0}^{\scriptscriptstyle DW}(s,t) = i A_{1}(\tilde{s},t) 
    {{\rho}^{\scriptscriptstyle DW}_{\lower 3pt \hbox{$\scriptstyle T$}}
    ({\bf q}_{\rm cm},{\bf k}_{\rm cm}) / q_{\rm cm}}\;, 
 \label{fdwia}
\end{equation}
%%%
where the distorted-wave tensor density has been introduced 
%%%
\begin{mathletters}
 \begin{eqnarray}
    &&{\rho}^{\scriptscriptstyle DW}_{\lower 3pt\hbox{$\scriptstyle T$}} 
    ({\bf q},{\bf k}) =
    \sum_{l=1}^{\infty} 
     \varrho_{\lower 3pt \hbox{$\scriptstyle l$}}(q,k)
     P'_{l}({\bf \hat{q}}\cdot{\bf \hat{k}}) \;, \\
     \varrho_{\lower 3pt \hbox{$\scriptstyle l$}}(q,k) &=&
     4\pi \int_{0}^{\infty} dr \, r^2
     \phi^{(+)}_{l q}(r)
     \rho_{\lower 3pt \hbox{$\scriptstyle T$}}(r)
     \Big[j_{\lower 2pt \hbox{$\scriptstyle l-1$}}(kr)+  
          j_{\lower 2pt \hbox{$\scriptstyle l+1$}}(kr)\Big]\;.
 \label{rhodw}
 \end{eqnarray}
\end{mathletters}
%%%
Note that $\phi^{(+)}_{lq}(r)$ is the $\eta$-meson distorted wave
with angular-momentum $l$ and $P'_{l}$ is the derivative of the 
Legendre polynomial of order $l$. In particular, this expression 
shows that there is no s-wave re-scattering for the $\eta$-meson, 
as $P'_{0}(x) \equiv 0$.

\section{Results}
\label{sec:results}
 
 In Fig.~\ref{figone} we display the coherent $\eta$-photoproduction
cross section from ${}^{40}$Ca at a photon laboratory energy of 
$E_{\gamma}=625$~MeV. In order to identify those effects arising
exclusively from relativity and final-state interactions we have 
factored out the elementary amplitude from the cross section. 
Thus, the only dynamical information that this calculation is 
sensitive to is nuclear structure and distortions. The dotted
line represents a plane-wave-impulse-approximation (PWIA) 
calculation in which the lower components of the single-particle
wave functions were determined from the free-space relation; the 
upper component remained unchanged, apart from a small normalization 
correction. This represents our best attempt at reproducing
standard nonrelativistic calculations which employ free, on-shell
spinors to effect the nonrelativistic reduction of the elementary 
amplitude. Indeed, in this ``nonrelativistic'' limit there is a
simple relation between the tensor and vector densities of 
closed-shell nuclei:
%%%
\begin{equation}
  \rho_{\lower 3pt \hbox{$\scriptstyle T$}}(Q) \approx
  -{Q \over 2M_{N}} 
  \rho_{\lower 3pt \hbox{$\scriptstyle V$}}(Q) \;,
 \label{trhotv}
\end{equation}
%%%
where $M_{N}$ is the free nucleon mass and  
$\rho_{\lower 3pt \hbox{$\scriptstyle V$}}(Q)$ is the
Fourier transform of the ground-state matter density.
In this way, the PWIA calculation establishes a
baseline, against which possible relativistic effects may be 
inferred. The effect of distortions on this nonrelativistic-like
calculation is depicted with the dashed line. At these low
energies ($q_{\rm cm}\alt 300$~MeV) the real part of the 
optical potential is attractive. This creates a competition 
with the imaginary part that results in a 
distorted-wave-impulse-approximation (DWIA) cross section 
relatively close to its plane-wave value. This picture, however,
changes dramatically once the lower components of the wave functions 
are determined dynamically, rather than from the free-space relation. 
In the particular case of the Walecka model the lower components are 
enhanced substantially in the medium as a consequence of the large 
scalar and vector mean-field potentials. Since the coherent cross 
section becomes dominated by the tensor density---which is very 
sensitive to this enhancement---we obtain a 
relativistic-plane-wave-impulse-approximation (RPWIA) cross
section that is, at least, twice as large as its nonrelativistic
counterpart (dot-dashed line). Note that as in the nonrelativistic 
case, distortion effects do not seem to play an important role at
this low energy (solid line). 

  In Fig.~\ref{figtwo}a) we show the corresponding behavior of the
coherent cross section at a photon energy of $E_{\gamma}=700$~MeV.
At these energies ($q_{\rm cm}\agt 400$~MeV) the real part of the
optical potential has become repulsive. This, in combination with
a relatively large imaginary component, results in a large 
quenching of the distorted-wave cross sections relative to their 
plane-wave values. Note that in spite of distortions---which make 
the interior of the nucleus, and thus the region of small effective 
nucleon mass, largely inaccessible---the relativistic calculation 
(RDWIA) still shows a substantial enhancement relative to the 
nonrelativistic result (DWIA). We also display the coherent
cross section---including the effect from the elementary 
amplitude---in Fig.~\ref{figtwo}b). We conclude, because the 
elementary amplitude varies relatively slow in this region, that 
most of the qualitative features identified in Fig.~\ref{figtwo}a) 
remain. Finally, in Fig.~\ref{figthree} we show a breakdown of the 
elementary contributions to the RPWIA calculation of 
Fig.~\ref{figtwo}b). This figure shows that a significant portion 
of the strength arises from the individual contributions from
the $D_{13}$(1520) excitation and the t-channel exchange of vector 
mesons, while as expected very little strength is contributed by 
$S_{11}$(1535) excitation or the Born terms. Also note that the 
constructive interference between $D_{13}$(1520) excitation and 
vector-meson exchange results in a cross section substantially 
stronger than their incoherent sum.

\section{Conclusions}
\label{sec:concl}
 We have computed the coherent $\eta$-photoproduction cross section
in a relativistic-impulse-approximation approach. Our formalism
differs from conventional approaches in that we do not perform a
nonrelativistic reduction of the elementary amplitude. In this 
manner, we could explore possible relativistic corrections to the 
coherent amplitude arising from medium modifications to the nucleon 
spinors. We have adopted a relativistic parameterization of the
elementary amplitude in terms of four Lorentz- and gauge-invariant 
amplitudes. Using general arguments we have shown that the coherent 
reaction is sensitive to only one of the components of the
elementary amplitude. Moreover, we identified the ground-state
tensor density---and not the vector density---as the essential
component of the nuclear structure. 

 The very high accuracy proton data~\cite{krus95a} and the inferred 
neutron data from the deuteron target experiment~\cite{krus95b} are 
well reproduced within the effective Lagrangian model which in turn 
is used to construct the elementary transition operator; this is the 
input to our nuclear calculations. The near cancellation of the 
sum of the proton and neutron $S_{11}(1535)$ helicity amplitudes allow
us to examine the background contributions in great detail. In
particular, our calculations indicate that the vector mesons and the 
$D_{13}(1520)$ account for most of the strength in the differential 
cross section. This can provide further theoretical insights into 
the background mechanism of eta photoproduction. For the nuclear 
tensor density we used a mean-field approximation to the Walecka model. 
The Walecka model is characterized by large scalar and vector 
potentials that induce a large enhancement in the lower 
(``small'') components of the single-particle wave functions. 
The tensor density depends linearly on these lower components 
and is, thus, sensitive to their in-medium enhancement. Indeed, 
we reported relativistic cross sections that were substantially 
larger than to those obtained from assuming the free 
lower-to-upper ratio, as is usually done in nonrelativistic 
calculations. Further, we demonstrated that in the low-energy 
region, where the real part of the optical potential is attractive, 
these large enhancements are preserved even in the presence of 
distortions.

 To conclude, we address two possible complications to the simple
picture presented here: 1) violations to the impulse approximation 
and 2) off-shell ambiguities in the elementary amplitude. In the impulse
approximation one assumes that the elementary amplitude---which only
contains on-shell information---can be used without modification in
the nuclear medium. However, any microscopic model of the reaction is
bound to predict some sort of modification to the elementary amplitude
as the process is embedded in the medium.  For example, the coupling
to intermediate $N^{*}$ resonances represents an important component
of most realistic models of the photoproduction amplitude. However, a
variety of processes, such as the interaction of the resonances with
the nuclear mean fields as well as Pauli blocking, can affect the
formation, propagation, and decay of these resonances in the nuclear
medium. Thus, it is important to have a reliable microscopic model in
which to test the validity of the impulse approximation. A microscopic
model can also provide guidance on how to take the elementary
amplitude off-shell. The form of the elementary amplitude used here,
although standard, is not unique. Many other choices---all of them
equivalent on shell---are possible. While all these choices are
guaranteed to give identical results for on-shell observables, they
can yield vastly different predictions off-shell. Without theoretical
guidance, there is no hope of resolving the off-shell ambiguity. 
Without experimental support, there is no way of testing the validity
of our models. Indeed, much work remains to be done---on both
theoretical and experimental fronts---before a clear picture of the 
coherent process can emerge.

\acknowledgments
This work was supported in part by the U.S. Department of Energy
under Contracts Nos. DE-FC05-85ER250000 (JP), DE-FG05-92ER40750 (JP), 
by the U.S. National Science Foundation (AJS), and by the Natural 
Sciences and Engineering Research Council of Canada (MB).

%%%%%%%%%%%%%%%%%%%%% Figures %%%%%%%%%%%%%%%%%%%%
\begin{figure}
 \caption{A comparison of the $p( \gamma , \eta )p$ and
          $n( \gamma , \eta )n$ elementary amplitudes to a sample of the
	  data from Ref.'s [13] (proton) and [14] (extracted neutron).
	  Calculations are shown for the full amplitude (Born terms, 
	  vector-meson exchange, and intermediate $S_{11}$(1535) and
	  $D_{13}$(1520) resonance excitation), as well as for the
	  intermediate $S_{11}$(1535) alone.}
 \label{figzero}
\end{figure}
%%%
\begin{figure}
 \caption{The coherent $\eta$-photoproduction cross section
	  from ${}^{40}$Ca at a photon laboratory energy of
	  $E_{\gamma}=625$~MeV; note that the elementary
	  amplitude $A_{1}$ has been factored out. 
	  The dotted (dashed) line represents a plane-wave
	  (distorted-wave) calculation in which the lower
	  components of the wave functions were determined 
          from the free-space relation. The dash-dotted
	  (solid) line represents a plane-wave
	  (distorted-wave) calculation in which the lower
	  components were generated dynamically from the
	  Dirac equation.}
 \label{figone}
\end{figure}
%%%
\begin{figure}
 \caption{The coherent $\eta$-photoproduction cross section
	  from ${}^{40}$Ca at a photon laboratory energy of
	  $E_{\gamma}=700$~MeV; with [a)] and without [b)]
          the elementary amplitude $A_{1}$ factored out. 
	  The dotted (dashed) line represents a plane-wave
	  (distorted-wave) calculation in which the lower
	  components of the wave functions were determined 
          from the free-space relation. The dash-dotted
	  (solid) line represents a plane-wave
	  (distorted-wave) calculation in which the lower
	  components were generated dynamically from the
	  Dirac equation.}
 \label{figtwo}
\end{figure}
%%%
\begin{figure}
 \caption{Breakdown of the elementary contributions to the
          coherent $\eta$-photoproduction cross section
	  from ${}^{40}$Ca at a photon laboratory energy of
	  $E_{\gamma}=700$~MeV. All curves were generated 
          in a relativistic plane-wave-impulse approximation.}
 \label{figthree}
\end{figure}
%%%%%%%%%%%%%%%%%%%%% Figures %%%%%%%%%%%%%%%%%%%%


\begin{references}
\bibitem{bofmir86}  S. Boffi and R. Mirando,
                    Nucl.~Phys.~{\bf A448}, 637 (1986).
\bibitem{cek87}     A.A. Chumbalov, R.A. Eramzhyan, and S.S. Kamalov,
                    Z. Phys. {\bf A328}, 195 (1987).
\bibitem{bentan90}  C. Bennhold and H. Tanabe,
                    Phys. Lett. B~{\bf 243}, 13 (1990);
                    Nucl.~Phys.~{\bf A530}, 625 (1991).
\bibitem{ndu91}     A. Nagl, V. Devanathan, and H. \"Uberall,
		    {\it Nuclear Pion Photoproduction,}
		    (Springer-Verlag, Berlin Heidelberg, 1991)	
\bibitem{tryfik94}  V.A. Tryasuchev and A.I. Fiks,
	            Phys. Atom. Nucl. {\bf 58}, 1168 (1995).
\bibitem{tieger84}  D.R. Tieger, E.C. Booth, J.P. Miller,
		    B.L. Roberts, J. Comuzzi, G.W. Dodson,
		    S. Gilad, and R.P. Redwine, Phys. Rev.
		    Lett.~{\bf 53}, 755 (1984).
\bibitem{gothe95}   R.W. Gothe, W. Lang, S. Klein, B. Schoch,
		    V. Metag, H. Str\"oher, S.J. Hall, and
		    R.O. Owens, Phys. Lett. B~{\bf 355}, 59 (1995)
\bibitem{ahrens93}  J. Ahrens et al., Photoproduction of $\eta$-mesons 
		    on ${}^{4}$He; MAMI-A2 and TAPS collaboration,
		    MAMI experiment A2/12-93.
\bibitem{cgln57}    G.F. Chew, M.L. Goldberger, F.E. Low, and Y. Nambu,
                    Phys. Rev. {\bf 106}, 1345 (1957).
\bibitem{serwal86}  J.D.~Walecka, Ann. of Phys. {\bf 83}, 491 (1974);
                    B.D. Serot and J.D. Walecka, Adv. in Nucl. Phys. 
                    {\bf 16}, J.W. Negele and E. Vogt, eds. 
                    (Plenum, N.Y. 1986).
\bibitem{benm92}    M. Benmerrouche, Ph.D. thesis, Rensselaer
		    Polytechnic Institute, 1992.
\bibitem{bmz95}     M. Benmerrouche, J.-F. Zhang and N.C.Mukhopadhyay,
	            Phys. Rev. D~{\bf 51}, 3237 (1995); 
                    N.C. Mukhopadhyay, J.-F. Zhang and
	            M. Benmerrouche, Phys. Lett. B~{\bf 364}, 1 (1995).
\bibitem{krus95a}   B. Krusche et al., 
	            Phys. Rev. Lett.~{\bf 74}, 3736 (1995).
\bibitem{krus95b}   B. Krusche et al., 
	            Phys. Lett. B~{\bf 358}, 40 (1995)
	  	    and private communication.
\end{references}
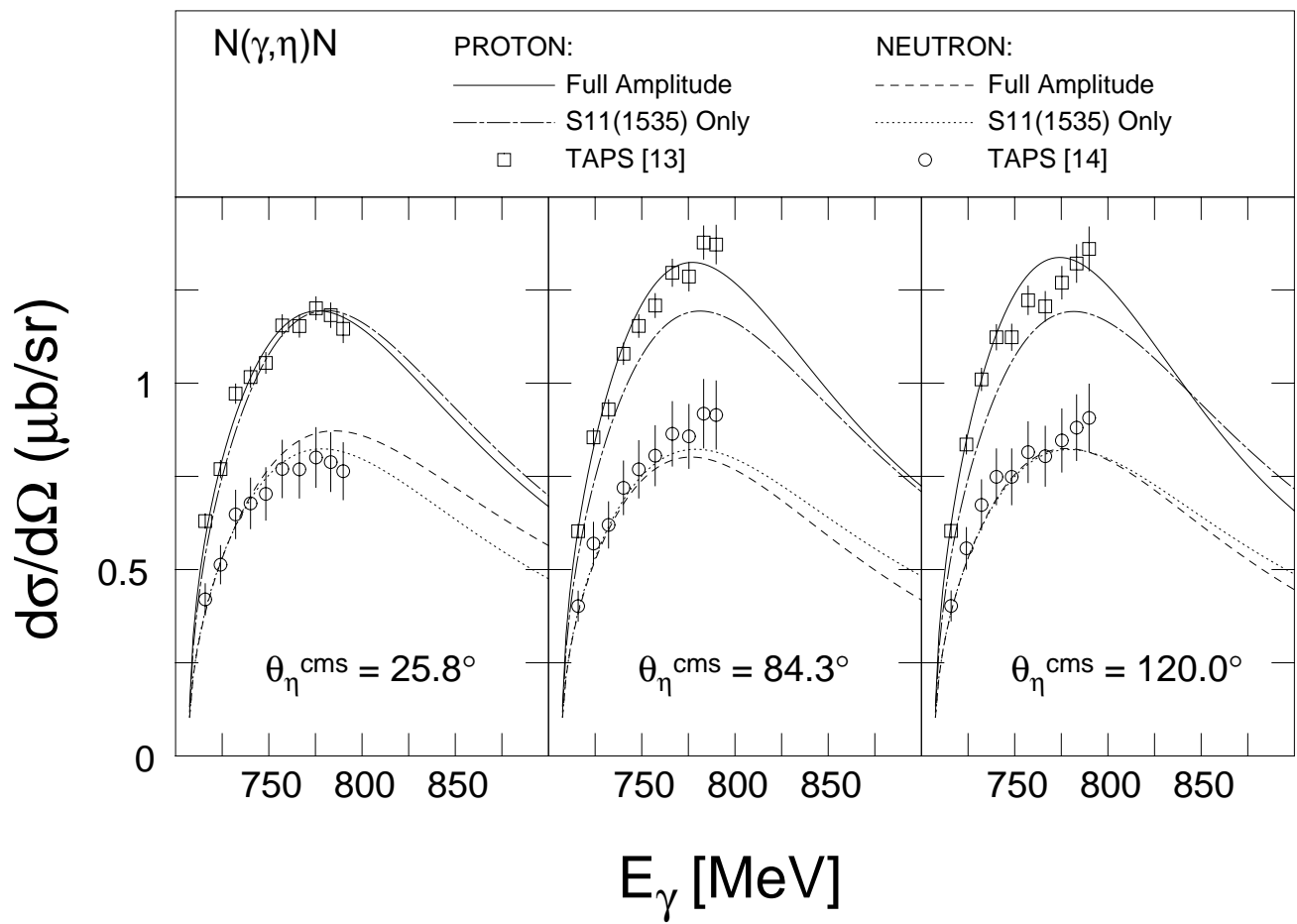
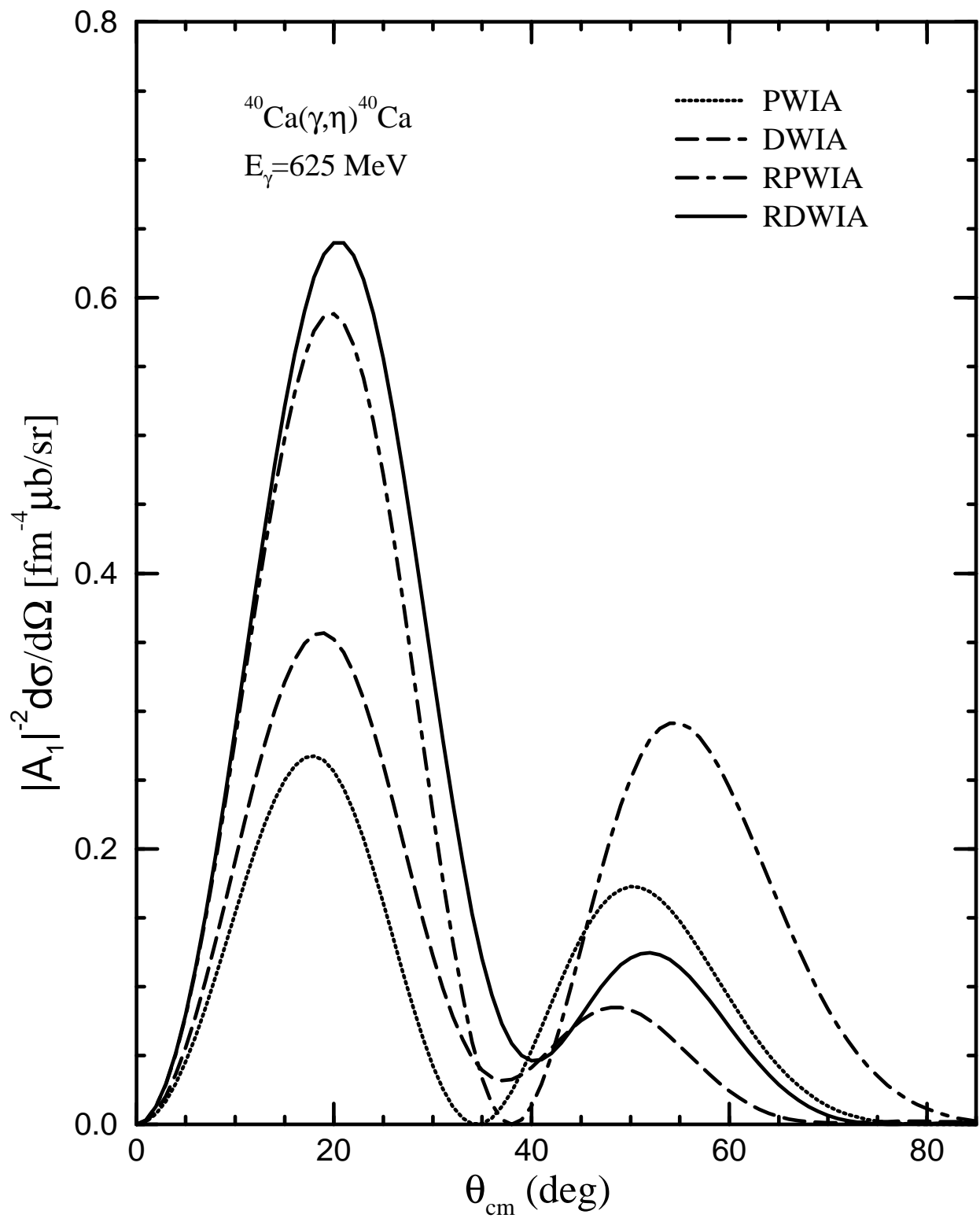
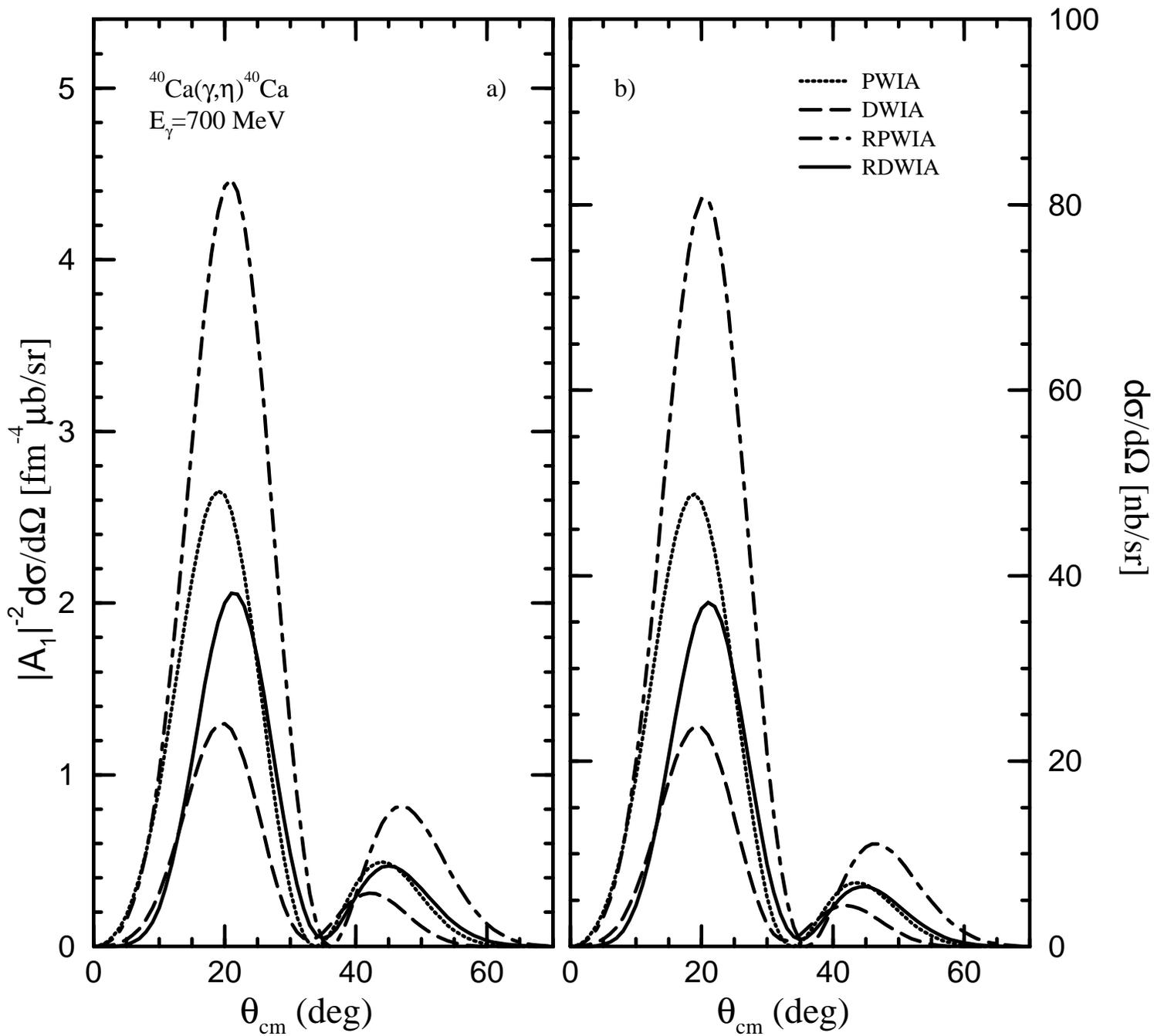
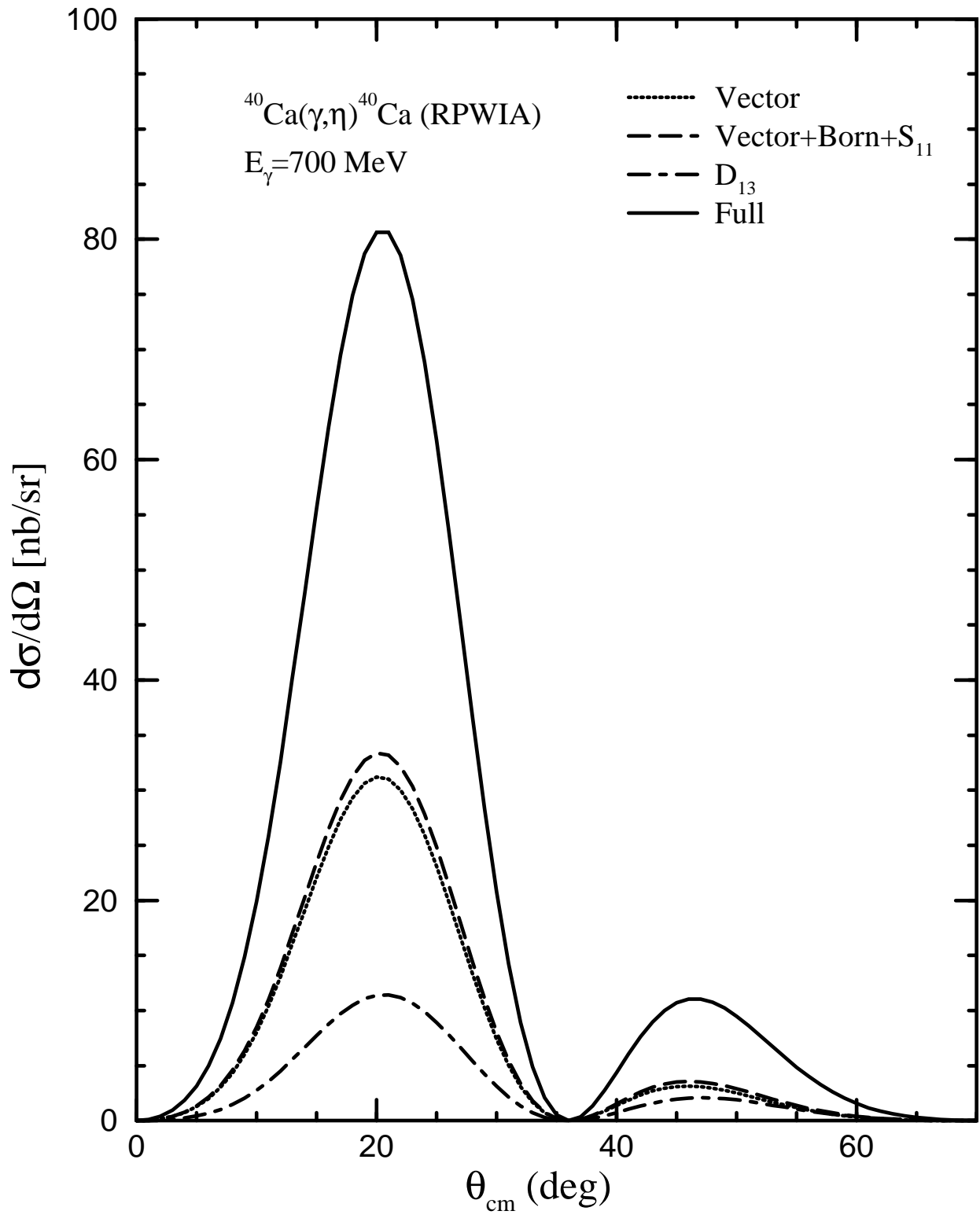
\end{document}